# Security and Communication Networks

# LogKernel: A Threat Hunting Approach Based on Behaviour Provenance Graph and Graph Kernel Clustering

Jiawei Li,[1] Ru Zhang,[1] Jianyi Liu,[1] and Gongshen Liu,[2]

[1] *Beijing University of Posts and Telecommunications, Beijing 100876, China.*
[2] *Shanghai Jiao Tong University, Shanghai 200240, China.*

Correspondence should be addressed to Ru Zhang; zhangru@bupt.edu.cn

## Abstract

Cyber threat hunting is a proactive search process for hidden threats in the organization's information system. It is a crucial component of active defense against advanced persistent threats (APTs). However, most of the current threat hunting methods rely on Cyber Threat Intelligence(CTI), which can find known attacks but cannot find unknown attacks that have not been disclosed by CTI. In this paper, we propose LogKernel, a threat hunting method based on graph kernel clustering which can effectively separates attack behaviour from benign activities. LogKernel first abstracts system audit logs into Behaviour Provenance Graphs (BPGs), and then clusters graphs by embedding them into a continuous space using a graph kernel. In particular, we design a new graph kernel clustering method based on the characteristics of BPGs, which can capture structure information and rich label information of the BPGs. To reduce false positives, LogKernel further quantifies the threat of abnormal behaviour. We evaluate LogKernel on the malicious dataset which includes seven simulated attack scenarios and the DAPRA CADETS dataset which includes four attack scenarios. The result shows that LogKernel can hunt all attack scenarios among them, and compared to the state-of-the-art methods, it can find unknown attacks.

## 1. Introduction

Advanced Persistent Threats (APTs) have the characteristics of persistence and concealment. These threats can bypass the Threat Detection Software (TDS) and lurk, so the enterprises information system may contain attacks that have already occurred but have not been detected. To better prevent and respond to such attacks, Endpoint Detection and Response (EDR) tools are widely deployed for enterprise security. However, these tools rely on matching low-level Indicators of Compromises (IOCs), which leads to 'alarm fatigue' problem and failure to reveal the complete attack scenario. To overcome this challenge, recent research solutions hunt for cyber threats by perform a causality analysis on audit logs [1,2,3,4]. In fact, causality and contextual information in audit logs imply high-level behaviours and goals of attackers that are difficult to hide.

Threat hunting is a proactive search process for latent attacks, which has become a key component of mitigating APT attacks. Existing work extracts attack behaviors from threat intelligence and designs matching algorithms to search for these known attacks from audit logs. To achieve this, some work [1,5] constructed audit logs as provenance graphs that contain rich contextual information and model threat hunting as a graph matching problem. Besides, THREATRAPTOR [4] designs Threat Behavior Query Language (TBQL) to query audit logs stored in the database. However, both graph matching and querying with TBQL need to extract attack behaviors from threat intelligence. These extracted attack behaviors are structured as attack graphs, in which nodes represent IOCs and edges represent IOC relations. These methods rely heavily on threat intelligence, which leads to some limitations. On the one hand, when there are deviations between threat intelligence and facts, attack activities may be missed. On the other hand, the description of the same APT attack event may come from different reports, so the information in these reports may be different or contradictory. Furthermore, attacks in threat intelligence are not comprehensive. Many APT attacks have not been disclosed by threat intelligence, and APT groups will upgrade cyber weapons or change intrusion strategies when attacking new targets. We refer to the above two cases as unknown attacks, and existing methods cannot detect these attacks.

Some solutions of investigating attack, such as matching rule knowledge base [2,6] or employing tag strategy [7,8], require manual participation of domain experts. And the completeness and accuracy of the expert knowledge will affect the analysis result. To overcome the above problem, Nodoze[9] builds an Event Frequency Database to replace the rule knowledge base, which considers that audit events related to attacks occur infrequently. However, to avoid detection, attackers will disguise themselves as normal behaviors or use some normal processes, such as *svchost.exe*, which affect the accuracy of methods for calculating threat scores based on matching or single event frequency.

To solve these problems in threat hunting, this paper proposes LogKernel, a threat hunting approach based on graph kernel clustering. It does not require additional expert knowledge to evaluate the threat in provenance graphs, nor does it require knowledge in threat intelligence to search for threat behaviors. Our first key insight is that there is a considerable discrepancy between the behavior of attackers and normal users, which is intuitively reflected by the disparity between the topological structure in provenance graphs. Therefore, LogKernel first abstracts the audit log into *Behavior Provenance Graphs* (BPGs) which can represent different behaviors. Then, it uses a graph kernel method to calculate the similarity between BPGs. However, the off-the-shelf graph kernel methods cannot be used directly, since the BPG is a kind of labelled directed graph with multiple types of directed edges. Therefore, we propose the BPG kernel, which is improved by the Weisfeiler-Lehman kernel and can capture the graph topology and the rich label information. Based on the calculated kernel values which indicates the degree of similarity, we cluster the BPGs by a clustering algorithm. Our second key insight is that the frequency of benign behaviors is much higher than that of threat behaviors, providing a basis for determining which clusters represent threats. Therefore, LogKernel determines which clusters represent threats based on the number of similar behaviors. Finally, considering false positives caused by low-frequency normal behaviors, a threat quantification method is proposed to estimate the degree of BPGs.

We evaluate the effectiveness and accuracy of LogKernel using three different datasets. In the first malicious dataset, we simulate seven attack scenarios, three from open APT reports and two complex attacks based on current attack techniques and strategies. In addition, we also execute two cyber weapons. Then we use the DAPRA CADETS dataset released by the Transparent Computing program to evaluate the applicability of LogKernel. Finally, we evaluate the false positives of our method on benign dataset that do not contain attacks and verify that the threat quantification method has a good performance in reducing false positives. The results show that LogKernel can hunt the attack scenarios effectively, and compared to the state-of-the-art methods relaying on Cyber Threat Intelligence (CTI), it can find unknown attacks.

In summary, this paper makes the following contributions:

1. LogKernel is proposed in this paper, which is a system that hunts for threats in organization's information systems. The system does not require additional expert knowledge and manual participation of domain experts. Meanwhile, it finds threats by comparing differences between behaviors, rather than matching similar attacks in logs with attack behaviors extracted from threat intelligence. Compared to most threat-hunting methods, LogKernel can hunt unknown attacks .

2. A novel Behavior Provenance Graph abstract algorithm is designed to construct BPGs from audit logs. In order to better represent the behavior characteristics, nodes and directed edges in BPGs are assigned labels based on the attributes of logs. Additionally, a novel density-based partitioning method is proposed to mitigate the impact of dependency explosion.

3. This is the first time that a graph kernel clustering method is proposed for threat hunting. In this paper, a novel graph kernel clustering approach are presented to measure the similarity between BPGs and cluster similar BPGs into one group. The graph kernel method is designed based on the WL kernel and message passing ideas, which can capture the topological structure of BPGs and quantify the similarity between BPGs efficiently and accurately.

4. We design a process to evaluate the effectiveness and accuracy of LogKernel. The performance of each stage is analyzed on several datasets and compared with other methods. The results show that Logkernel can hunt threats like other work. Furthermore, it can hunt unknown threats as well.

## 2. Related Work

### 2.1. Threat Hunting

***Provenance analysis.*** Our work relays some ideas in provenance analysis, so we introduce prior work in this area. The idea of constructing a provenance graph from kernel audit logs was introduced by King et al[10]. Then, the provenance graph is widely used for threat hunting, attack detection [11,12], attack investigation [13,14,15] and scenario reconstruction [16,17]. All these researches encountered a variety of challenges.

For attack detection, the main challenge is that the graph size grows continuously as APTs slowly penetrate a system. UNICORN [12] uses a graph sketching technique to summarize long-running system execution to combat slow-acting attacks. Attack investigation mainly traces the root causes and ramifications of an attack through causal analysis. OmegaLog[14] bridges the semantic gap between system and application logging contexts, and merges application event logs with system logs to generate the universal provenance graph(UPG).The challenges of attack scenario reconstruction is the semantic gap between the low-level logs and the attack behavior. Holmes [2] maps low-level audit logs to TTPs and APT stages by matching the rules in knowledge base. WATSON [3] infers log semantics through contextual information and com-bines event semantics as the representation of behaviors. It also can reduce analysis workload by two orders of magnitude for attack investigation. Nevertheless, the amount of audit logs generated by a typical system is nontrivial, limiting the efficiency of log analysis and resulting in dependency explosion [19]. To solve this problem, resent researches have proposed execution unit partition [20,21], taint propagation [22,21], grammatical inference [23,18], and universal provenance [14] techniques. These researches can perform more accurate provenance tracking and reduce storage and time overhead. The scope of LogKernel differs from these researches, since we intend to hunt unknown threats via the comparison between the BPGs.

***Threat hunting.*** Threat hunting is becoming an essential element of active defence against advanced persistent threats. POIROT [1] constructs query graph by extracting IOCs together with the relationships among them from CTI reports. Then, the query graph is used to match the most similar subgraph in the provenance graph. Its core contribution is the implementation of the above process via an in-exact graph matching algorithm. However, the query graph of POIROT requires time-consuming manual construction by cyber analysts. THREATRAPTOR [4] provides an unsupervised, light-weight, and accurate NLP pipeline that can extract structured threat behaviors from unstructured OS-CTI texts. It designs Threat Behavior Query Language (TBQL) to facilitate threat hunting in system audit log data.

Extracting threat knowledge from CTI reports and matching them in audit logs can facilitate threat hunting. However, the above researches can only hunt known attacks disclosed by CTI reports. In fact, there are still some APT attacks that have not been discovered or disclosed. To hunt these unknown threats, a threat hunting method based on graph kernel clustering is proposed, which can hunt for unknown threats without knowledge of CTI reports.

## 2.2. Graph kernel

Graph kernels are kernel functions that compute the inner product of a graph [24], which can be intuitively understood as functions measuring the similarity of pairs of graphs. The graph kernel [25] approach was proposed by Thomas Gaertner for graph comparison. There is a huge amount of work in this area due to the prevalence of graph-structured data and the empirical success of kernel-based classification algorithms [26].

Compared with traditional machine learning methods, graph kernels allows kernelized learning algorithms such as support vector machines to work directly on graphs without extraction to transform them into fixed-length, real-valued feature vectors, which loses a lot of structured information. In this paper, a new graph kernel method is proposed, which can capture structure information and rich label information of labeled directed graphs.

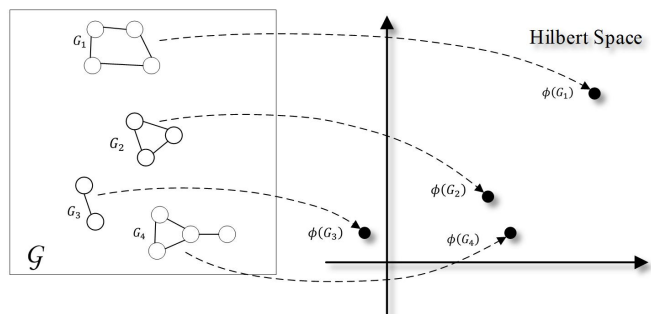

Figure 1: Illustration of the kernel-based graph mapping

Let $\mathcal{G}$ be a graph set and $k: \mathcal{G} \times \mathcal{G} \rightarrow \mathbb{R}$ be a function associated with a Hilbert space $\mathcal{H}$, so there exists a map $\phi: \mathcal{G} \rightarrow \mathcal{H}$ with $k(G_1, G_2) = \langle \phi(G_1), \phi(G_2) \rangle$ for all $G_1, G_2 \in \mathcal{G}$. Then $\langle \cdot, \cdot \rangle$ denotes the inner product of $\mathcal{H}$ and $k$ is said to be a positive definite kernel function. Figure 1 shows a graph kernel implicitly mapping graphs to a Hilbert space $\mathcal{H}$.

Given a graph $G = (V, E, L)$, where $V$ denotes the set of nodes, $E$ denotes the set of edges and $L$ denotes the set of labels of nodes. For a node $v$, we define neighbourhood $N(v) = \{v' | (v, v') \in E\}$ to denote the set of nodes to which $v$ is connected by an edge, and then $|\mathcal{N}(v)| = deg(v)$ is the degree of node $v$.

The most important and well-known of these strategies is the Weisfeiler-Lehman (WL) algorithm and kernel [27]. In this paper, WL subtree graph kernel is selected and improved to calculate the kernel values between the nodes of the graphs. The basic idea of kernel calculation is as follows:

First, we assign an initial label $l^{(0)}(v)$ to each node of $G_1$ and $G_2$. In labeled graph this label is $l_i \in L$ and in unlabeled graph it is a degree, i.e. $l^{(0)}(v) = deg(v)$.

Next, we iteratively assign a new label to each node base on the current labels within the node's neighborhood:

$$l^i(v) = relabel((l^{i-1}(v), sort(\{\{l^{i-1}(u) | u \in \mathcal{N}(v)\}\}))) \tag{1}$$

Where the double-braces are used to denote a multi-set. $sort(S)$ realizes sorting the elements in multiple-set in ascending order, and then $l^{i-1}(v)$ is added to the front of the set. $relabel(S)$ maps $S$ to a new label which has not been used in previous iterations.

After running K iterations of re-labeling, we now have a label $l^i(v)$ for each node that summarizes the structure of its K-hop neighborhood. Then we can summary statistics over these labels and calculate the kernel values.

Message Passing Graph Kernels [28] is a graph kernel framework which consists of two components. The first component is a kernel between vertices and the second component is a kernel between graphs [16]. Let $k_v$ be a kernel between nodes and $k_\mathcal{N}$ be a kernel between neighborhoods. Then, compute the kernel $k_v^t$ between each pair of vertices iteratively and the recurrence is shown below:

$$k_v^{t+1}(v_1, v_2) = \alpha k_v^t(v_1, v_2) + \beta k_\mathcal{N}(\mathcal{N}(v_1), \mathcal{N}(v_2)) \tag{2}$$

Where α and β are nonnegative constants. After computing the kernel between each pair of vertices for *T* iterations, we can compute a kernel between graphs as follows:

$$k_G(G_1 G_2) = k_V(V_1 V_2) \tag{3}$$

In this paper, we propose a BPG kernel on the framework to calculate the similarity between BPGs. Specifically, we improve the WL kernel based on the characteristics of BPGs for calculating the kernel values between nodes. Then, we compute the $k_G(G_1 G_2)$ and get a positive definite kernel matrix $K_{n \times n}$, which can be considered as a similarity matrix in a Hilbert space.

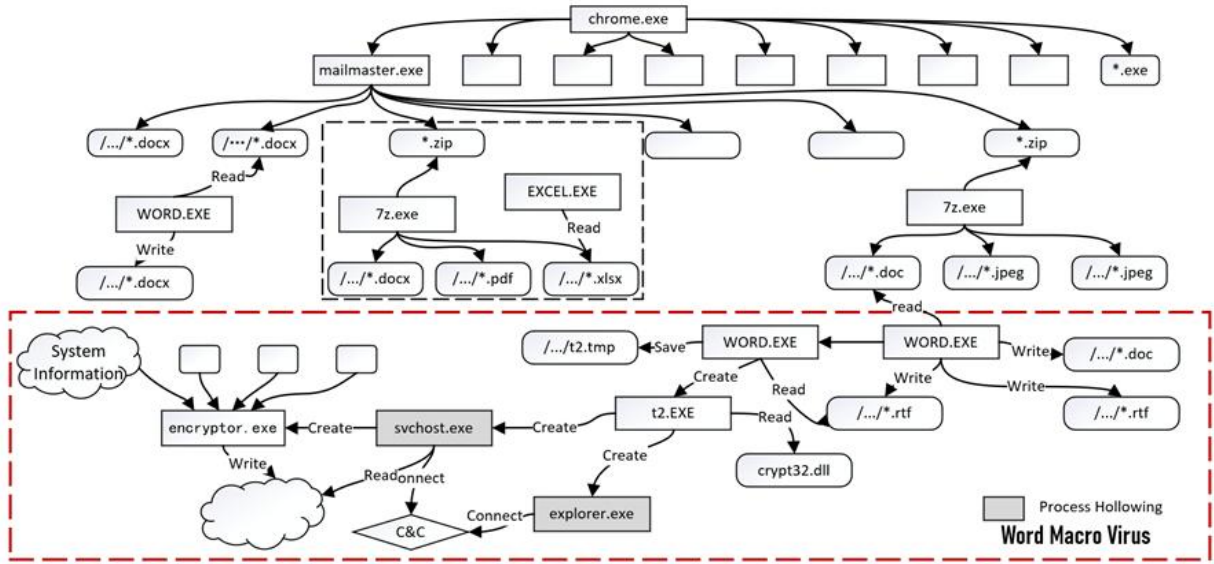

Figure 2: The provenance graph for the macro viruses attack scenario. Rectangles represent processes, diamonds represent IPs, and rounded rectangles represent files

## 3. Motivation

In this section, we introduce the motivation of the approach. We first use an attack example to illustrate the challenges and limitations of the threat hunting based on provenance analysis, and then analyse the feasibility of our approach.

### 3.1. Limitations and Challenges

**An Attack Scenario.** Ever since macros were introduced, they have been maliciously exploited by hackers. More recently, macros have also been widely used as attack vectors by advanced Persistent Threat (APT) organizations [29]. To avoid detection, macro viruses use process hollowing [30]. Next, we will introduce an attack scenario using macro viruses.

Consider a scenario where a worker in an organization read emails and download attachments (such as *Office files* or *zip packages*) every morning. One of the emails is a phishing email with a zip file attached. The worker unzips the package and opens the document containing the macro instructions. At the same time, the macro virus begins to work. First, the virus releases a PE file named *t2.tmp* and executes it. Then the *tmp* file uses process hollowing techniques to inject malicious code into *explorer.exe* and *svchost.exe*, which tests the connection to the C&C servers. After that, *svchost.exe* queries the registry and collects host information. Finally, the process encrypts the information and sends it to the C&C server in POST mode.

System audit logs record the OS-level operation such as writing file, executing process and connecting IP address. Specifically, we can abstract a triple (*Subject, Object, Relation*) from audit logs, where Relation is an operation, Subject is the entity executing operation and Object is the entity being operated. These triples are used to build a provenance graph [9] for tracking information flows in audit logs. Figure 2 demonstrates the complete scenario represented by a provenance graph, with the red dotted boxes representing the execution of the macro virus and the black dotted boxes representing the normal behavior of the worker.

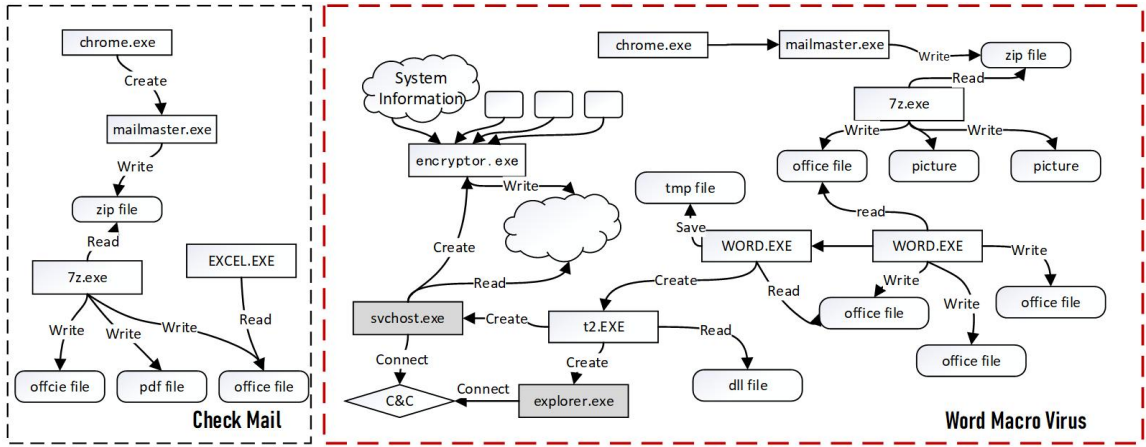

Figure 3: Two different Behavior Provenance Graphs. Check mail is A normal behavior and Word Macro Virus is a threat behavior

Based on the above attack scenario, We illustrate the limitations and challenges of existing threat hunting and attack investigation approach:

**Dependency Explosion.** Most existing approaches or systems based on provenance graph face the *dependency explosion problem.* The main reason for this is that some processes have a long lifetime and iterative input/output processes. For example, mailbox client process receives and sends a large number of emails during its lifetime [19]. The process is considered as a single node in the provenance graph, which results in threat behavior and normal behavior appearing in the same graph. As shown in Figure 2, the attack behavior in the red dotted box is constructed in one graph with normal behavior due to *mailmaster.exe*.

**Relay on Knowledge.** In essence, both the rule knowledge base and threat intelligence rely on knowledge to match the threat behavior in the provenance graphs. Unfortunately, some situations in practice can cause the threat path to break. For example, the rule matching methods miss the hollowed *svchost.exe* and *explorer.exe*, resulting in the threat path interruption. In addition, many audit behaviors of attackers are common in the audit logs of normal users, such as writing office files or connecting to the external network, and attackers use some techniques to disguise themselves to avoid detection. This condition leads to a low threat score calculated based on the frequency of a single event. Another obvious problem is that they can only hunt threats described by existing knowledge.

### 3.2. Feasibility Analysis

Two different behaviors described in Figure 3 intuitively reflect our first key insight. Checking email is a normal behavior that happens frequently. The topology of the BPG describing macro virus behavior varies substantially from normal behavior such as the label of nodes and the relationships between nodes. In fact, the topology implies high-level behaviors and goals. Because the purpose of the attacker is different from that of the normal user, the actions taken by the attacker and the causal relationship between the actions are also different. For example, the topology of sending a macro virus is similar to checking email, but the causal relationship between subsequent operations implicitly includes the attacker's purpose of disguising himself and stealing information. Thus, we can compare the similarity between the provenance graphs of different behaviors to separate advanced threats from benign activities, and model threat hunting as a clustering problem of labeled directed graphs.

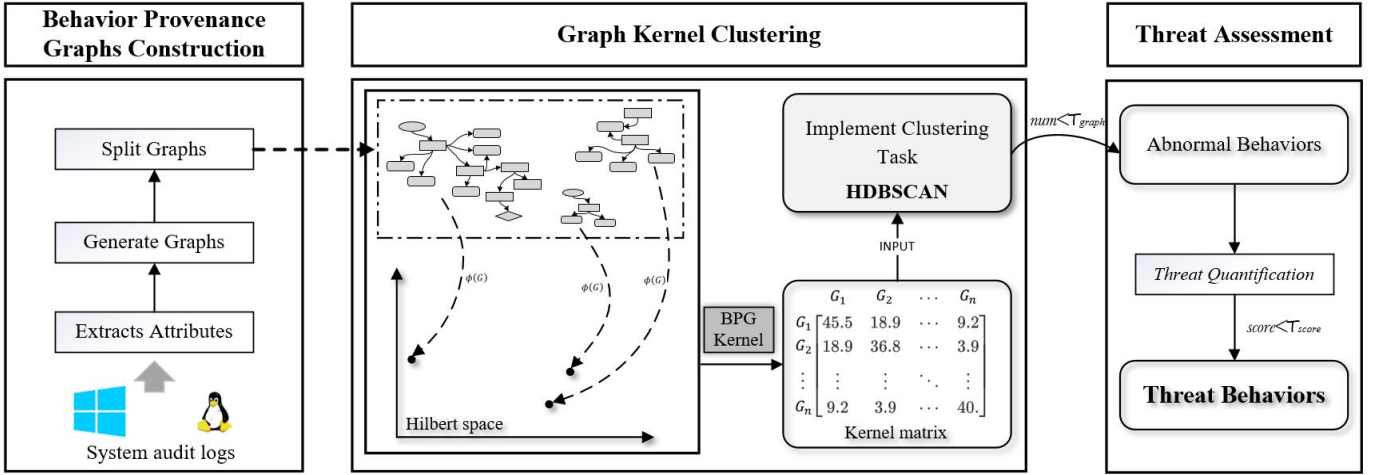

Figure 4: Overview of LogKernel

We design an algorithm in Section 4.1 to extract BPGs from audit logs to describe different behaviors. Then we choose the graph kernel to implement our graph clustering task, which is a popular method for measuring the similarity between graphs. However, the existing graph kernel cannot be directly used to perform the similarity measurement due to two reasons. First, the BPGs nodes not only have distinct types, but also contain attributes that are significant for determining similarity. Besides the type of edges is also a critical feature. Second, the graph kernel method is typically used for classification tasks and the data are balanced samples, while the number of graphs in each cluster of BPG clustering tasks vary greatly. An unsupervised graph kernel clustering approach is proposed in Section 4.2.

## 4. Approach

The overall approach of LogKernel is shown in Figure 4, which consists of three phases: BPGs construction, Graph Kernel Clustering and Threat assessment. First of all, LogKernel analyses audit logs for information on entity types, attributes and relationships between entities. According to the information, logs are constructed as BPGs, which describes various behaviors. Next, the similarity between the BPGs is calculated through an improved graph kernel, and the BPGs are clustered using a clustering algorithm. Finally, abnormal behavior is identified based on the frequency of similarity behavior. In order to reduce false positives, a threat quantification method is introduced to assess risk.

### 4.1 Behavior Provenance Graph Construction

The construction process of the Behavior Provenance Graph (BPG) is described in detail in this section. The BPG is proposed to represent different behaviors in this paper. Compared to the existing work [1-4], nodes and edges of BPG are assigned appropriate labels to represent the characteristics of behaviors. As discussed in Section 3.1, long-running processes cause false dependencies in the provenance graph, which causes different behaviors to appear in the same graph. So, a density-based partitioning method is proposed to remove false dependencies to generate more concise BPGs.

**Definitions 1.** *Behavior Provenance Graph* **(BPG)**: the labeled directed graph $G = (V, E, L_V, L_E)$ represents the Behavior Provenance Graph, where $V$ represents the node set of system entities, $L_V = \{l_{v_i} | v_i \in V\}$ indicates the label set of nodes. $E = \{\cup e_{ij}\}$ is the directed edge set, and $L_E = \{l_{e_{ij}}\}$ is a label set, where $l_{e_{ij}} = \{e_{ij}: Relationship\}$ denotes operation between entities. The entity types and relationships between entities in this work are shown in Table 1. In fact, a BPG describes the information flow of a certain behavior at the system level. For example, a BPG can represent the complete process of downloading an attachment in an email, modifying it, and forwarding it.

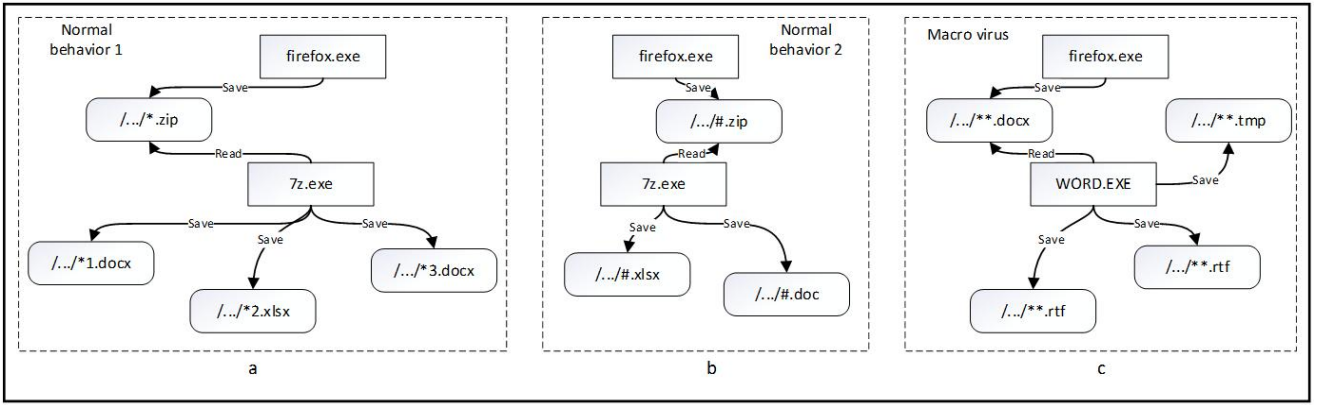

Figure 5: Subgraphs of BPGs to illustrate the importance of Node Label

Table 1: Entity types and relationships

| Start Node | End Node | Relationships |
|---|---|---|
| | File | Read; Write; Execute |
| Process | IP | Connect |
| | Process | Create |
| IP | User | Logon |
| User | Process | Execute |

**Definitions 2. Node Label:** Node labels are used to indicate the characteristics of behaviors. *Relationships* can be used directly as the label of the edge, such as $l_{e_{ij}} = \{e_{ij}: Read\}$. However, the entity type cannot be directly used as the node label, which will ignore the rich attribute information of the entity. Rich attribute information and relationships imply the purpose of behavior, but some information is not useful for similarity calculation. For example, the same malware executed in different hosts may have different paths and names of the generated files.

To measure the similarity between BPGs, the attribute information of the entity is mapped to labels and assigned to nodes. Function GETNODE in algorithm 1 abstracts attribute information as labels. For *process,* the label is process name, such as *mailmaster.exe* and *svchost.exe.* The label of the *file* only considers the file type and not the path information. For example, *report.doc* and *data.xls* in *D:\download\* are assigned the same label of *office file*. The label of *IP* is address and port, while the label of *User* is the user's name. Function GETEDGE abstracts the relation as edge and operation as the label of edge.

An example is presented below to illustrate the importance of Node Label. As shown in Figure 5, there are three subgraphs of BPGs. (a) and (b) denote behaviors that normal users download and unzip the zip files from the emails, and (c) denotes the behavior of macro virus releasing PE file. (a) and (c) are isomorphic when only node types with distinct forms are considered. Consider another scenario where the entity attributes are directly used as the label of the node, such as %name%.docx. In this case (a) and (b) will be considered as different behaviors due to the labels.

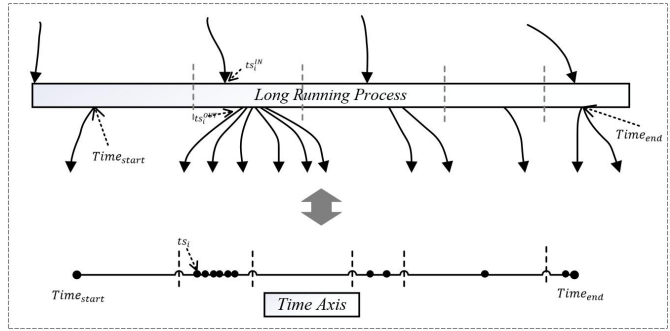

Figure 6: Density-based partitioning method. The dependencies are expressed as points on the time axis.

In order to extract BPGs that describe different behaviors, a forward depth-first search (DFS) is performed on the provenance graph. *File* type nodes are the starting point of DFS, because they only appear at the end of the directed edges, as illustrated in Table 1. In the process of traversing the graph, two operations are performed to prevent false dependencies. First, we compare the time of directed edges to prevent information flowing from a future event to a past event. Second, the long-running processes are partitioned to remove false dependencies and achieve the segmentation of the provenance graph.

To partition the long running process, density-based partitioning method is proposed. In other words, the long-running process is partitioned by the density of dependencies during its life cycle. A common phenomenon at the system level is that when a process has dependencies on multiple entities in a short period of time, these dependencies belong to the same behavior instance. The dependencies are expressed as points on the time axis for calculating their density, as shown in Figure 6. Note that the edges *from* and *to* node are separately calculated. $Time_{start}$ is the time when the first dependency appears, and $Time_{end}$ is the time when the last dependency appears. Let $ts_i$ indicate the timestamp of the occurrence of the *i-th* dependency, and calculate a time interval sequence $T = \{T_1, T_2, \ldots, T_n\}$, where $T_i = ts_{i+1} - ts_i$ denotes the time interval between adjacent dependencies. The formula for calculating the density of the *i-th* node is as follows:

$$Density_i = \frac{Time_{end} - Time_{start}}{T_{i-1} + T_i} \qquad (4)$$

When the density of the node is high, it suggests that there exist dependencies in a short period of time before and after this node. We traverse all the nodes and consider continuous dependencies with the density higher than the average density as belonging to the same behavior instance. In this process, the long running process is divided into several partition units.

The edges to node are also divided into different execution partitions through the above process. Ultimately, we judge which edges belong to the same behavior instance. To prevent information flowing from a future event to a past event, the time of edges to node should be before the edges from node, which is shown by $ts_i^{IN} < ts_j^{OUT}$. $ts_i^{IN}$ denotes the time when the last dependency occurred in the execution partition, and $ts_j^{OUT}$ denotes the time when the first dependency occurred in the execution partition.

| | |
|---|---|
| **Algorithm 1**: Graph Abstraction Algorithm | |
| **Input**：*OS-level logs* | |
| **Output**: *Behavior Dependency Graphs* | |
| 1 | *Logs* ← READLOGS(*OS-level logs*) |
| 2 | *Processes* ← LONGRUNNINGPROCESS(*OS-level logs*) |
| 3 | **for all** *log* in *Logs* **do** |
| 4 | *Type* = GETNODETYPE(*log*) |
| 5 | *NodeAttribute$_i$* = GETTYPENODE(*log*) |
| 6 | *NodeList*.append(**NodeAttribute$_i$**) |
| 7 | **end for** |
| 8 | *ProvenanceGraph* = GENERATEGRAPH(*NodeList*) |
| 9 | V, $L_V$ ← GETNODE (*NodeAttribute*) |
| 10 | E, $L_E$ ← GETEDGE (*NodeAttribute*) |
| 11 | *BehaviorGraphs* = SPLITGRAPH(*ProvenanceGraph, Processes*) |
| 12 | *Density* ← CALCULATEDESITY(*Processes*) |
| 13 | *Units* ← PARTITION(*Density*) |
| 14 | **return** *BehaviorGraphs* |

In order to construct BPGs, a graph abstraction algorithm is proposed. First, the algorithm obtains entity information from audit logs, as shown in Lines 3 through 7. The function GETNODETYPE determines the type of entity in the log, and then GETTYPENODE extracts attributes based on the entity type $NodeAttribute_{Processes} = \{ID, name, Target\_ID *, Opreation *\}$. *NodeList* stores the attributes of all entities for the construction of BPGs. Next, function GENERATEGRAPH generates provenance graph and assigns labels to nodes and edges. Finally, function SPLITGRAPH performs forward depth-first search (DFS) on the provenance graph and partitions the long-running processes which is described in the above section.

**4.2 Graph Kernel Clustering**

BPG kernel is proposed in this paper to calculate the similarity between behavior provenance graphs. It consists of two components. The first component is an improved WL kernel to calculate kernel values between nodes, and the second component is to calculate the kernel values between graphs.

The traditional WL graph kernel is only suitable for the undirected graph. An improved WL kernel is proposed to calculate kernel values between nodes in directed label graphs. The improved kernel only compares each node's multi-label set without label compression. Considering the type of directed edges, a kernel function for comparing directed edges is proposed. The kernel value is calculated by multiple iterations, which is more helpful to distinguish nodes in different BPGs. The improved WL kernel is described in the following.

First, according to definition 1, $G = (V, E, L_V, L_E)$ is a labeled directed graph. The elements in $L_V$ and $L_E$ are strings, which are not suitable for the process of multiset-label sorting and label compression in WL method. Therefore, we convert the string label sets into numeric label sets, in which each string element corresponds to a unique value. By this means, each node in $G$ is assigned an initial numerical label $l^{(0)}(v)$.

Next, we iteratively assign a new label to each node. The traditional WL graph kernel gets the multiset from the label set of neighborhoods, which does not consider the relationship between nodes. BPG kernel improves WL kernel by introducing multiset of edges. Given a node $v_i \in G$ and $\mathcal{N}(v_i) = \{u|(v_i, u) \in E\}$, where $\mathcal{N}(v_i)$ denotes the neighbour node set of $v_i$, and the multiset is $M(v_i) = \{(l_e(v_i, u), l^{i-1}(u))|u \in \mathcal{N}(v)\}$, where $l_e(v_i, u)$ is the numerical label of edge. Like the WL kernel, the elements of multiset are sorted in ascending order, and then $l^{(0)}(v)$ is added to the front of the set.

Then, we calculate the kernel values between nodes. It is necessary to take full account of the particularity of BPGs, that similar behavior graphs are not completely isomorphic. Therefore, set $v_1 \in G_1$ and $v_2 \in G_2$, $k_v^1(v_1, v_2)$ is calculated by comparing the label in $M(v_i)$, The calculation formula is as follows:

$$k_v^1(v_1, v_2) = |M(v_1) \cap M(v_2)| \tag{5}$$

Where $k_v^1(v_1, v_2)$ denotes the similarity of subgraphs with $high = 1$, which are generated by the nodes $v_1$ and $v_2$ as the root. Finally, we compute the kernel $k_v^T$ between each pair of vertices iteratively, and formula is as follows:

$$k_v^{t+1}(v_1, v_2) = \alpha\, k_v^t(v_1, v_2) + \beta k_{edge}(u_1, u_2) \cdot \sum_{u_1 \in \mathcal{N}(v_1)} \sum_{u_2 \in \mathcal{N}(v_2)} k_v^t(u_1, u_2) \tag{6}$$

Where $k_{edge}(u_1, u_2)$ is defined as the kernel value of the edge, which means that if the label of two edges is the same, the kernel value is assigned 1, otherwise it is 0. The formula is as follows:

$$k_{edge}(u_1, u_2) = \begin{cases} 1, & if\ l_e(v_1, u_1) = l_e(v_2, u_2), \\ 0, & otherwise. \end{cases} \tag{7}$$

In essence, the kernel value $k_v^T(v_1, v_2)$ is a quantification of the similarity of the subgraphs, in which the shortest step from $v_i$ to any node is less than T. In our method, the best effect is achieved when T=5.

The second component of the BPG kernel is to calculate the kernel values between the graphs based on the kernel value between nodes. First, we build a mapping set $B(V_1, V_2)$ between nodes based on the idea of optimal assignment kernel, which can provide a more valid notion of similarity [1]. $V_i$ is the node set of $G_i$, and the mapping is only operated between nodes of the same type. Then $B(V_1, V_2) = \{v_1: v_2 | k_v^T(v_1, v_2) = k_{max}\}$, where $k_{max} = \max\{k_v^T(v_1, v_2'), k_v^T(v_1', v_2)\}$ and $v_2' \in V_2$ has the same type as $v_1$. Note that the mapping is from a set with fewer nodes to a set with more nodes. The formula for calculating kernel values between BPGs is as follows:

$$k_G(G_1 G_2) = \sum_{(v_1, v_2) \in B} k_v^T(v_1, v_2) \tag{8}$$

Finally, a positive definite kernel matrix $K_{N \times N}$ is calculated by the BPG kernel, where $K_{i,j}$ is the kernel value between $G_i$ and $G_j$. Kernel matrix can be considered as a similarity matrix which represents pair-wise similarities (inner products) between $G_i$ and $G_j$ in a Hilbert space.

Since the number of types of behavior abstracted from the logs is unknown, supervised learning methods (SVM, etc.) are not suitable for our task. In this paper, a typical clustering method called HDBSCAN [31] is employed. Compared with other clustering methods, HDBSAN is more advisable for our data. The main reasons are as follows:1) it does not need to declare the number of clusters in advance, 2) It has good robustness to outliers, and 3) it supports input custom similarity(distance) matrix. In our approach, HDBSCAN code published by McInnes L et al. [32] is adopted to implement clustering task, which takes the kernel matrix generated by BPG kernel as input.

**4.3 Threat Assessment**

After clustering, the graphs representing attack behavior and normal behavior are divided into separate clusters. According to the second key insight above, LogKernel determines which clusters of graphs are abnormal. In order to reduce false positives, threat quantification method is performed on abnormal BPGs to find the threat behaviors.

***Determines abnormal behaviors.*** When the number of graphs in a cluster is less than the $Threshold_{graphs}$, the behaviors represented by BPGs in the cluster are determined to be abnormal behaviors. However, an obvious issue is that some of these abnormal behaviors are performed by normal users in low frequency, which leads to high false positives.

***Threat quantification method.*** To overcome the above problem, *Threat quantification method* is proposed to evaluate the threat level of abnormal BPGs. There are several characteristics of advanced cyber threats. For APT attacks, executing malicious files, collecting sensitive information and connecting to the C&C server are necessary operations [10]. Based on these characteristics, we design a quantification method. The execution of malware has been implicitly expressed by BPGs, so our method uses malicious domain name access, sensitive information acquisition and privilege escalation as the criteria for threat quantification.

First, the threat values of IP and URL are quantified through public databases and the frequency of appearances in audit logs. An online database of malicious IP [43] and Alexa Traffic Rank [44] are employed to identify unsafe web resources, and then some IP that cannot be determined are assigned scores by frequency. Secondly, different threat values are assigned to different types of sensitive information, such as account information, sensitive databases, and sensitive files. These are generally labeled within enterprises or automatically identified by tools [33]. Finally, it is a critical step for the attacker to elevate the user's privileges or login as a higher-privileged user such as root, which provides a prerequisite for the attacker's subsequent operations. The final threat quantification formula is as follows:

$$Threat\_Score = \sum_{i=1}^{k} \left( \alpha f_{ip}^i + \beta f_{user}^i + \gamma f_{sens}^i \right) \quad (9)$$

Where $f_{ip}^i$ denotes the threat values of IP and URL, $f_{user}^i$ represents the quantification of user permissions and $f_{sens}^i$ denotes the quantification of sensitive information. $\alpha, \beta$ and $\gamma$ represent the weight of three criteria, which can be adjusted according to actual requirements. Ultimately, we rank the threat score of the abnormal BPGs. If the threat score exceeds a threshold value, Logkernel determines that the BPGs represent threat behavior and raises an alarm.

# 5. EVALUATION

## 5.1 Experimental Datasets

We evaluate LogKernel's efficacy and accuracy on three datasets.1) A malicious dataset, which comes from the real work environment and contains 7 attack scenarios. 2) The DAPRA CADETS dataset, which is a public dataset. 3) A benign dataset, which comes from the real work environment without cyber-attacks.

Table 2: Attack scenarios in the malicious dataset.

| Attack Scenario | Description | Key nodes and operation |
|---|---|---|
| OceanLotus [34] | Using phishing mails to deliver a malicious payload and a malicious sample, and decrypt the sample to load additional data. Then releasing the white application file of the adobe reader. after loading it connects to C&C sever. | Node:{*hat* file, *%random%*.exe, *%deceive %*.docx，*dll* files, C2 sever}<br>Operation {**execute** hat file, **release** Malicious files and **deceive** document, **connect** C2} |
| APT28[35] | Using the macro file to release the Trojan file and modify the registry to realize self-starting after booting. Then encrypting the collected files and sending them back. | Node {*%macro%*.xls, *%Trojan%*.exe, *%malicious%*.dll, C2 sever}<br>Operation {**execute** macro, **decrypt** Trojan, **release** malicious *dll*, **connect** C2} |
| Kimsuky[36] | Using process injection to evade the intrusion detection system, then escalating privileges to obtain host information, and finally sending it to C&C sever. | Node {*%malicious%*.scr, registry, explorer.exe(*Process Hollowing*), privileges, C&C sever }<br>Operation {**execute** malicious scr, **write** registry, **inject** code to process, **connect** C2} |
| Unknown attack 1 | Using phishing emails to deliver macro virus samples, which release PE files and perform process hollowing. Finally encrypting the collected information and sending it to C&C sever (Attack Scenario) | Node {*%macro%*.doc, *%PE%*.tmp, explorer.exe&snchost.exe (*Process Hollowing*), C2 sever}<br>Operation {**download&execute** macro, process hollowing, **encrypted** information, **connect** C2} |
| Unknown attack 2 | Using weak passwords for remote login. Then getting higher privileges user information in the host and accessing the registry information. Finally, exfiltrating collected information over FTP to remote servers | Node {remote user, root, C2 sever}<br>Operation {**remote** login, **login** root, **encrypted** information, **connect** C2} |
| cyber weapons | Using two homologous cyber weapons with no initial intrusion and delivery phases, and seeing what happens when logs are incomplete | |

*Malicious dataset.* In order to obtain the dataset containing attacks, we added some hosts and virtual machines to the existing work environment and collected audit logs from all hosts in the work environment. Like in previous work [1-5], we simulated seven attack scenarios on these hosts and virtual machines, as shown in Table 2. Among them, three real attack scenarios come from the public APT report, two synthetic APT scenarios are designed based on the attack methods and strategies in the public report, and also two homologous cyber weapons from third-party releases are used [34]. To simulate real attack scenarios, we set the attack time span to 7 days and continuously executed extensive ordinary user behaviors and underlying system activities in parallel to the attacks on the added hosts and virtual machines. We used ETW to collect logs in Windows system, and use camflow [42] to collect logs in Linux system.

In the malicious dataset, in addition to attack behaviors, the normal behaviors mainly include remote login, mail checking, code modification and execution, and software installation, etc.

***DAPRA CADETS.*** DAPRA CADETS dataset [38] is released by the DARPA Transparent Computing program, which contains APT attacks. The dataset was collected from hosts during DARPA's two-week red team vs. blue team Engagement 3 in April 2018 [39]. In this engagement, some normal operations such as SSH login, Web browsing and email checking will be executed on the hosts. At the same time, attackers will use various APT attack method to penetrate the system and steal privacy information. The attacks on CADETS were executed four times, resulting in 44,404,339 system level audit entries.

Table 3: Attribute information of the above three datasets

| **Datasets** | **Size** | **logs** | graphs | |
| --- | --- | --- | --- | --- |
| | | | *Attack* | *Benign* |
| Malicious dataset | 10.3GB | 8,796,458 | 7 | 1867 |
| CADETS | 35.7GB | 44,404,339 | 4 | 1683 |
| Benign dataset | 15.6GB | 13,446,341 | 0 | 2453 |

Table 4: Clustering result of some scenarios

| **Scenario** | **Min Distance** | **number of graphs** | **accuracy** |
| --- | --- | --- | --- |
| OceanLotus[35] | 11.468 | 1 | 100% |
| APT28[36] | 10.734 | 1 | 100% |
| Kimsuky[37] | 9.278 | 1 | 100% |
| Unknown attack 1 | 3.136 | 1 | 100% |
| Unknown attack 2 | 5.121 | 1 | 100% |
| Cyber weapons | 4.257 | 2 | 100% |
| Check mails | 0.594 | 483 | 72.3% |

***Benign dataset***. Benign dataset comes from the real work environment. We collect system audit logs of common user behaviors and low-level service activities during two weeks. The main behaviors performed by these users include connecting SSH, executing code, browsing web sites, and downloading files. To better verify the applicability of the method, we collected audit logs from two GPU servers at the same time, including the behavior of ordinary users and administrators. The attributes of the above three datasets are shown in Table 3.

In all the experiments, the values of $Threshold_{graphs}$ and $Threshold_{Score}$ are set to 3 and 3600 respectively. This choice is described in detail in section 5.3. The performance is evaluated on the malicious dataset and CADETS dataset. The threshold values are validated on benign dataset.

**5.2 Experimental Result**

In this section, the performance of LogKernel is evaluated by a large number of experiments on the malicious dataset and CADETS dataset. First, the accuracy of graph kernel clustering is analyzed and visualized. Next, we evaluate the threat quantification method to demonstrate its effectiveness in eliminating false positives. Then, the accuracy of hunting threats was analyzed. And we conduct several sets of comparative experiments to illustrate the effectiveness of the BPG and BPG kernel proposed in this paper in threat hunting. Finally, we analyse and discuss the advantages of LogKernel compared with other methods.

*5.2.1 accuracy of graph kernel clustering*

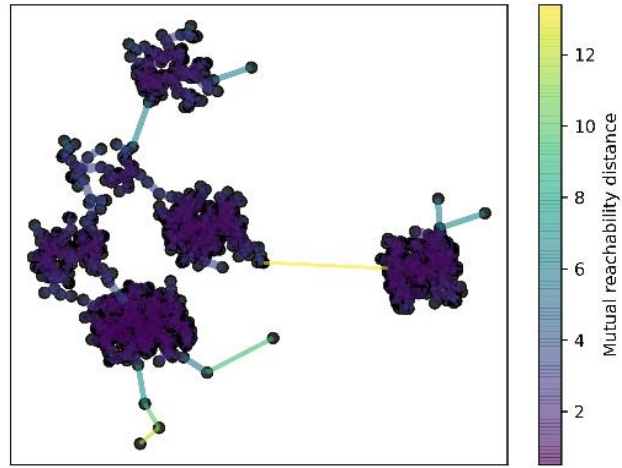

Figure 7: 2D visualization of mutual reachability distance and outliers indicate threat BPGs.

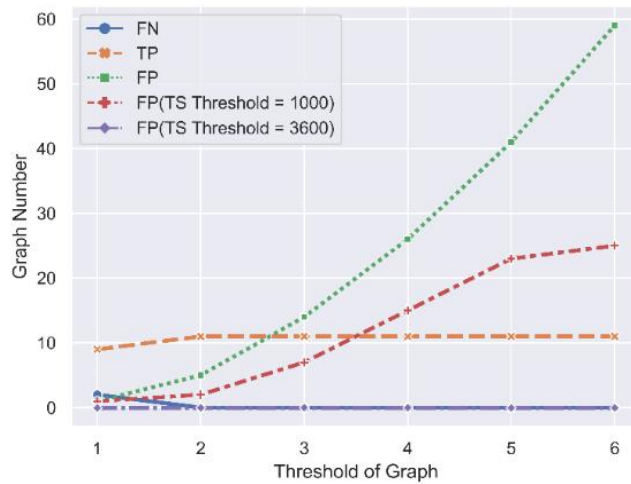

Figure 8: The influence of different thresholds on the results.

The accuracy of the graph kernel clustering indicates whether the attack behaviors are successfully separated from the normal behaviors. After the BPGs are abstracted, the threat BPGs are marked, which is used to track intermediate results. At the same time, the normal BPGs of checking mails are also marked. Figure 7 shows the 2D visualization of mutual reachability distance and outliers indicate threat BPGs. Table 4 shows the clustering result, where *Min Distance* represents the minimum distance from other clusters, and Number of Graphs indicates the number of BPGs that denote the attack scenarios and behaviors. For example, the closest dot to Kimsuky is Unknown Attack 1, as shown in Figure 7. The reason is that they both use process hollowing techniques for *explorer.exe* and the collected host information is encrypted and sent back to the C2 server. In the Behavior Provenance Graphs of the two scenarios, the subgraphs describing the above behaviors are similar, so the distance between the two graphs is relatively close. However, there are also many dissimilar parts in these two graphs, so they are not clustering in the same cluster.

Checking mails behaviors are further analyzed. Due to the differences in subsequent operations caused by different attachment types, the BPGs containing mail-related processes are divided into multiple clusters. However, some BPGs that do not contain mail processes appear in these clusters, along with checking mail BPGs in other clusters. The reason is that the behavior for some attachments is more similar to the behavior in other class clusters.

Finally, the accuracy of the check mails is 72.3%. The above situation does not affect our final results. This paper focuses on clustering the threat BPGs into the correct clusters, and the results show that the clustering for the threat BPGs is 100%.

*5.2.2 Evaluate Threat Quantification Method*

To evaluate the performance of threat quantification methods in reducing false positives, we conduct experimental evaluations on the malicious dataset and CADETS dataset. In the scenarios of using and not using the threat quantification approach, we count the number of false negatives, true negatives and false positives in the result. Furthermore, $Threshold_{graphs}$ is also critical to avoid false signals. We select six values (1-6) to evaluate the impact on the result.

The result is depicted in Figure 8. When the number of graphs in the cluster does not exceed the threshold of graph, these graphs are determined to be abnormal behaviors. As the threshold increases, the number of false alarms also increases. When using the threat quantification method, the number of false alarms is significantly reduced. The number of false alarms can be reduced to 0 by selecting an appropriate threat score threshold. This demonstrates that the use of the threat quantification method in LogKernel can effectively reduce false alarms caused by low-frequency abnormal behaviors.

*5.2.3 Accuracy of Hunting Threat*

Then the accuracy of the LogKernel is analyzed, and Table 5 shows the hunting results on two datasets. It can be seen that LogKernel can hunt all attack scenarios without false positives. In the malicious dataset, five complete attack scenarios appeared in five different clusters, and two homologous cyber weapons were clustered in the same cluster. Among these clusters, the maximum number of graphs is 2 which is lower than the $Threshold$ value of 3. However, there are multiple clusters where the number of graphs is not greater than 3. The graphs in these class clusters represent behaviors that users rarely perform. For example, in order to install a program on the mobile phone, a worker accesses the relevant file from social software to the mobile phone. The above behavior appears only twice in the malicious dataset. We perform threat quantification on these anomalous behaviors and find that the threat values of non-threatening behaviors are all below the threshold.

The threat quantification of the above graphs shows that the threat scores of all threat BPGs exceed the threshold, while those of normal low-frequency behaviors had threat values well below the threshold. Finally, LogKernel successfully hunts all threats without false positives on malicious dataset and CADETS dataset.

The effectiveness of LogKernel is evaluated by several comparative experiments that mainly contain the following two parts. First, the existing graph kernel methods is compared with BPG kernel proposed in this paper. Note that these graph kernel algorithms are not specifically designed for BPGs, and we use the public codes [40,27,28] to calculate the similarity between BPGs. Second, we consider two other cases of abstracting graphs and compare them with BPGs to illustrate the importance of abstracting attribute information as Node Label which is introduced in section 4.1. The first is no label graphs which hide the node labels of BPGs and only considering the type of nodes. The second is attribute label graphs. Instead of mapping the attribute information to Node Label, we use the attribute information directly as the label of the nodes to get the attribute label graphs. For example, *C:\Windows\System32\%name%.dll* represents the attribute information of a file object.

The results of comparative experiments are shown in Table 5. When only the node type is considered, false negatives are generated. When the logs are constructed as attribute label graphs, it results in a lot of false positives in our approach. Besides, the traditional WL kernel and MPGK AA are compared with BPG kernel. The number of iterations of the two kernel is set 5 which can put the best results. MPGK AA employs the theory of valid optimal assignment kernels for developing kernel based on Message Passing Graph Kernels, and its code is published on GitHub [41]. As shown in Table 5, the performance of LogKernel outperforms MPGK AA and WL kernel.

Table 5: Hunting results on Malicious dataset and CADETS dataset

| Dataset | Graph Kernel | Graph type | Recall | Precision | F-score |
|---|---|---|---|---|---|
| Malicious dataset | BPG Kernel | BPGs | 100% | 100% | 100% |
| | | No label graphs | 57.1% | 57.1% | 72.7% |
| | | Attribute label graphs | 100% | 43.8% | 60.9% |
| | WL Kernel[39,26] | BPGs | 85.7% | 50% | 63.2% |
| | MPGK AA[27] | | 42.9% | 30% | 35.3% |
| CADETS dataset | BPG Kernel | BPGs | 100% | 100% | 100% |
| | | No label graphs | 25% | 20% | 22.2% |
| | | Attribute label graphs | 100% | 50% | 66% |
| | WL Kernel | BPGs | 75% | 42.9% | 54.6% |
| | MPGK AA | | 50% | 33.3% | 40% |

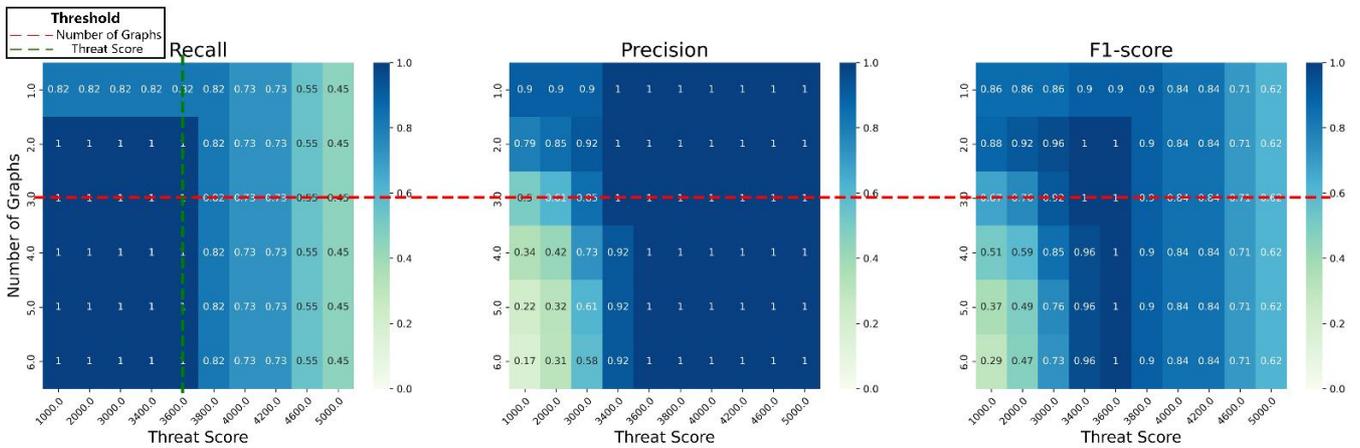

Figure 9: Determine the optimal threshold values.

### 5.2.4 Approach Comparisons

Compared with the state-of-the-art approaches (e.g. Poirot [1] and THREATRAPTOR[3]), LogKernel can hunt known attacks, such as three attack scenarios simulated from the APT reports and four attacks in CADETS. Additionally, LogKernel can also hunt for unknown attacks that have not been disclosed by threat intelligence. For example, unknown attack 1 is designed by us in combination with the attack characteristics of multiple APT organizations. Since this attack scenario does not appear in threat intelligence, current methods based on threat intelligence cannot hunt this attack. We used the attack scenario to represent the upgrading of attack technology by APT organizations and the attacks that have not been discovered or disclosed.

Compared with WATSON[3], which can abstract high-level behaviors from audit logs and reduce the analytical workload of attack investigations, LogKernel focuses on comparing and clustering behaviors to realize threat detection. WATSON's purpose is to abstract high-level behavioral and semantic information from contextual information in audit logs. The method applies heuristics to specify system entities as termination conditions for DFS during the extraction subgraph phase, which causes the complete behavior to be split into multiple subgraphs.

### 5.3 False positive analysis

***Determine the Optimal Threshold Value.*** The selection of the threshold value is critical to reducing false alarms. For example, too low the threat score threshold could cause some normal behaviors to be misclassified as attack behaviors, while too high the threshold could result in false negatives.

Thus, there is a trade-off in choosing an optimal threshold value. To determine the optimal threshold value, we measured recall, precision and F1-score using varying threshold values, as shown in Figure 9. When the $Threshold_{graphs} = 3$ and the $Threshold_{graphs} \in [3387, 3743]$, the F1-score, the harmonic mean of precision and recall, is at its peak. In fact, 3387 is the maximum score of the normal Behavior Provenance Graph, and 3743 is the minimum score of the threat Behavior Provenance Graph. Another reason we set the $Threshold_{graphs}$ to 3 is to consider the homology of malware, although this is not the scope of our work. When executing homologous malwares, attack Behavior Provenance Graphs with similar topological structures will be generated, which means when homologous malware is discovered in the information system, it will be classified into the same cluster. Therefore, we set the $Threshold_{graphs} = 3$ and $Threshold_{Score} = 3600$ as the optimal threshold values.

***Evaluation on Benign Dataset.*** We use the benign dataset to validate the threshold values and LogKernel. The benign dataset comes from the real work environment. Despite the fact that cyber-attacks are not included in the dataset, there are some normal behaviors that do not occur frequently. Besides, workers are inquisitive to open some domains that are not frequently accessed, or use higher-privileged users to access secret information or execute processes during work. As a result, we execute LogKernel on the benign dataset. Inevitably, LogKernel recognizes some abnormal behaviors that occur infrequently. However, when we quantify the threat of these abnormal BPGs, the highest threat score is 2980, well below the threshold. Compared to real attacks, these abnormal behaviors have a shorter path and only involve several types of untrusted IPs, user privilege escalation, and sensitive information, but not all of them. Therefore, their threat score cannot reach the threshold.

### 5.4 SYSTEM PERFORMANCE

To measure the performance of LogKernel, we record the running overhead of the system on malicious datasets and DAPRA CADETS. The scale and magnitude of these two datasets are similar to user data within an organization or enterprise. The runtime overhead of the system is divided into two parts: the first part is the overhead of reading all the audit logs from disk and generating the BPGs, and the second part is the overhead of finding the threat behavior from the BPGs. We perform the experiments on a server with an Intel(R) Xeon(R) Silver 4215R CPU (with 8 cores and 3.20 GHz of speed each) and 256 GB of memory running on Ubuntu 18.04.5 LTS.

Table 6 LogKernel performance overhead

| Datasets | Attack cases | Size on Disk | BPGs construction time | Graphs Size | | | Search Time |
|---|---|---|---|---|---|---|---|
| | | | | Nodes | Edges | Size | |
| DAPRA CADETS | cadets_1 | 11.1 GB | 25min28s | 133.1K | 295.6K | 57.6MB | 134.24s |
| | cadets_2 | 17.7 GB | 40min53s | 171.1K | 408.3K | 78.5MB | |
| | cadets_3 | 6.77 GB | 19min47s | 94.9K | 171.9K | 35.2 MB | |
| Malicious dataset | | 10.3GB | 22min19s | 106.2K | 183.6K | 42.9MB | 94.62s |

**BPGs construction.** In DAPRA CADETS, attacks are carried out in three different time periods, so the BPGs were constructed from three segments that are divided according to the attack cases, as shown in Table 6. Attacks simulated in the malicious dataset overlap in time, so they are not segmented by attack cases. Table 6 shows LogKernel performance overhead. The third column shows the initial size of the logs on disk, and the fourth column represents the time it takes to read the audit logs from disk into memory and construct the BPGs. The runtime overhead of BPGs construction depends on the number of audit logs and the operating system. In addition, the current experiment uses a single host, and the efficiency of the system can be further improved by parallel processing of multiple hosts. The fifth column represents the size of the constructed graph, including the number of nodes and edges and the size on the hard disk.

**Threat searching.** The sixth column represents the time spent hunting threats on both datasets. This time includes graph kernel clustering and threat quantification. In the hunting stage, we combined all the graphs of the three cases in the CADETS dataset for search. The results in the table show that analysing the BPGs constructed by 34.5GB large-scale log data and finding all the attacks in it can be completed in a relatively short time (134.24s). Comparing LogKernel with POIROT and THREATRAPTOR's fuzzy search mode, the total running time of these three systems is of the same order of magnitude. And LogKernel is faster than POIROT when processing the same size of data.

## 5.4 CASE STUDY

In order to understand the hunting process of LogKernel more intuitively, we consider Kimsuky[36] as a case study from Table 5. For the attack scenario, we manually analyze the accuracy of BPGs and evaluate the performance of LogKernel.

In this case, the infection starts with a classic executable file with *scr* extension, which is used by Windows to identify Screensaver artifacts. A worker intends to implement a function in the project, so he downloads similar code from the public website for reference, but this zip file contains malware *scr* file. It writes a *dll* file and sets the registry key to gain persistence. Then the explorer.exe injection performs by the *dll* file to avoid Anti-Virus detection. Finally, the malware contacts the C&C sever and sends back the encrypted information about the compromised machine. At the same time, the worker also downloads documentation, data samples, and installs Python.

As shown in the black dashed box in Figure 10, the complete provenance graph is generated and *chrome.exe* is a long running process causing false dependencies between multiple behavior instances. LogKernel's graph abstraction algorithm successfully separates these behavioral instances according to the density and occurrence time of the related dependencies. A normal *exe* file appears in the Kimsuky BPG because it occurs close to the download of the zip file, but its impact is almost negligible. The PDF and CSV files come from the same website, and the worker downloads both files at about the same time, so they are considered to belong to the same BPG. After a long time, the worker downloaded and installed Python from the official website, which is a new behavior instance.

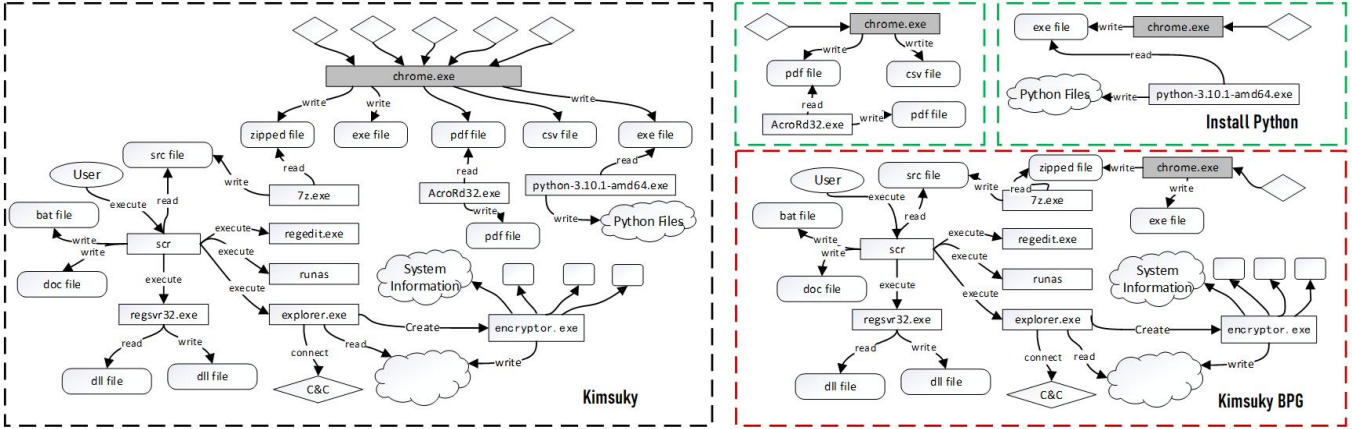

Figure 10: Case study to analyze the accuracy of BPGs and evaluate the performance of LogKernel

Then LogKernel computes the kernel values between these BPGs and clusters them. As shown in Figure 10, the topological structure of Kimsuky BPG is quite different from that of normal BPGs, which leads to the low kernel value between them. After clustering, there are many behaviors of downloading pdf and csv files in a certain cluster. However, since Python was installed only twice during the experiment, it was judged to be abnormal behavior.

Finally, LogKernel quantifies Kimsuky and Installing Python BPGs to achieve threat hunting. Because installing Python does not involve these three operations, the threat score is well below the threshold. In Kimsuky BPG, the attacker host information contains some labeled sensitive information, which makes the threat score higher than the threshold. Even if we design new methods of leaking data, such as using a network disk which does not appear in the threat report, LogKernel can still find the unknown threat.

## 6. DISCUSSION & LIMITATIONS

In this section, some limitations and possible extensions of LogKernel will be discussed.

The basic assumption of our method is that audit logs are trusted and cannot be tampered or destroyed. In fact, this is an important assumption for almost the entirety of recent work in provenance-based forensic analysis. Ensuring the integrity of audit logs is beyond the scope of this work. In addition, LogKernel cannot detect attacks that do not use system call interfaces because these cannot be captured by the underlying provenance tracker. But such behaviors appear to be rare, and the harm they can bring to the rest of the system is limited. Finally, LogKernel cannot hunt the attacks exploiting OS kernel vulnerabilities.

For LogKernel, the definition of labels requires some manual involvement. For common file types, we can use the extension to automatically assign the file's label, such as *.zip*, *.rar* are marked as zipped file. For some uncommon file types, manual participation of experts is required to assign appropriate label to such files. In fact, after manually labeling this type of file, automatic labeling can be performed later.

The density-based partitioning method in this paper is less accurate than the existing execution partitioning system [19,20,21]. These systems require complex binary program analysis to instrument a target application for execution partitioning at runtime [15]. During the experiment, there will be some normal behavior instances that are not separated from the threat behavior, such as the exe file in the case study. However, the errors produced by our method do not have a decisive impact on the results.

LogKernel is an offline system that analyses audit logs of Windows and Linux systems. It still scales well for other formats of logging. Extra work is simply to extract entities and relationships.

## 7. CONCLUSION

In order to reduce the dependence on additional expert knowledge and hunt unknown attacks, this paper proposes a threat hunting approach based on graph kernel clustering, which consists of three parts: BPGs construction, Graph Kernel Clustering and Threat Assessment. First, a graph abstraction algorithm is proposed to construct Behavior Provenance Graphs (BPGs), in which a density-based partitioning method is proposed to alleviate the dependency explosion problem. Second, based on the characteristics of the BPGs, a BPG kernel is proposed which can capture structure information and rich label information, and then HDBSCAN is used to perform the clustering task. Finally, LogKernel determines which clusters are abnormal, and the threat quantification method is performed on abnormal BPGs to hunt the threat behaviors. Experiments are carried out on three datasets containing simulated APT attacks, unknown attacks, public attack datasets and a large number of normal behaviors. Experimental results show that our method can hunt all threats in the dataset, and can detect unknown attacks compared with the method based on threat intelligence.

### Data Availability

The data DAPRA CADETS supporting this paper are from previously reported studies and datasets, which have been cited. These prior studies (and datasets) are cited at relevant places within the text as references [38, 39]. The processed data are available from the corresponding author upon request.

The other data used to support the findings of this study are available from the corresponding author upon request.

### Conflicts of Interest

The authors declare that they have no conflicts of interest.

### Acknowledgments

This work is supported by National Natural Science Foundation of China U1936216, U21B2020, and Fundamental Research Funds for the Central Universities (Beijing university of posts and telecommunications) for Action Plan under Grant 2021XD-A11-1.